\documentclass[twocolumn, aps, pre, superscriptaddress, showpacs, floatfix, noshowkeys, numerical]{revtex4-1}
\usepackage{dcolumn}
\usepackage{bm}
\usepackage{textcomp}
\usepackage{subfigure}
\usepackage{amsmath, amsfonts, amssymb, hyperref}
\usepackage{graphicx}

\newcommand{\td}[1]{\frac{{d}{#1}}{{d}t}}
\newcommand{\pd}[2]{\frac{\partial{#1}}{\partial{#2}}}

\newcommand{\bp}[1]{\left({#1}\right)}
\newcommand{\bc}[1]{\left\{{#1}\right\}}
\newcommand{\bs}[1]{\left[{#1}\right]}

\newcommand{\bb}{\begin{eqnarray}}
\newcommand{\ee}{\end{eqnarray}}
\newcommand{\nn}{\nonumber}

\begin{document}

\title{Emergence of skew distributions in controlled growth processes}

\author{Segun Goh}
\affiliation{Department of Physics and Astronomy and Center for Theoretical Physics, Seoul National University, Seoul 151-747, Korea}

\author{H. W. Kwon}
\affiliation{Department of Physics and Astronomy and Center for Theoretical Physics, Seoul National University, Seoul 151-747, Korea}

\author{M. Y. Choi}
\email{mychoi@snu.ac.kr}
\affiliation{Department of Physics and Astronomy and Center for Theoretical Physics, Seoul National University, Seoul 151-747, Korea}

\author{J.-Y. Fortin}
\affiliation{Institut Jean Lamour, D\'epartement de Physique de la Mati\`ere et des Mat\'eriaux, Nancy Universit\'e, BP 70239 F-54506 Vandoeuvre les Nancy Cedex, France}

\begin{abstract}
Starting from a master equation, we derive the evolution equation for the size distribution of elements in an evolving system, where each element can grow, divide into two, and produce new elements.  We then probe general solutions of the evolution equation, to obtain such skew distributions as power-law, log-normal, and Weibull distributions, depending on the growth or division and production. Specifically, repeated production of elements of uniform size leads to power-law distributions, whereas production of elements with the size distributed according to the current distribution as well as no production of new elements results in log-normal distributions.  Finally, division into two, or binary fission, bears Weibull distributions. Numerical simulations are also carried out, confirming the validity of the obtained solutions.
\end{abstract}

\pacs{05.40.-a, 89.75.Fb, 05.65.+b}

\maketitle

\section{Introduction}

The central limit theorem is one of the most important theorems in probability theory. It states that the sum of a large number of independent random variables has a Gaussian distribution, which is symmetric. In reality, however, there exist many phenomena to which the central limit theorem is not applicable. In most of those cases, asymmetric skew distributions arise, characterized by long tails on one side; prototypic examples include power-law, log-normal, and Weibull distributions.
In contrast to the Gaussian distribution, of which the origin is well known, there still lacks appropriate understanding of the origin of skew distributions.

The power-law distribution is observed in various physical, biological, economical, or social systems and there are extensive studies attempting to understand its properties \cite{Lucchin85,Levy97,Gabaix03};
\textit{inter alia}, scale invariance is the most important property.
While Zipf's law \cite{Zipf49} and Pareto's law \cite{Pareto96} provide well-known examples of the power-law distribution, the Yule process \cite{Yule25}, Gibrat's law \cite{Gibrat31}, or preferential attachment \cite{Barabasi99} has been proposed as the mechanism for the power-law distribution.
Related to the power-law distribution, the log-normal distribution has been used as an alternative of the former \cite{Fleming72, Oexle98, Jo07}. Unlike the Gaussian distribution, the log-normal distribution emerges from multiplication (rather than addition) of a large number of independent random variables \cite{Limpert01}.
Thus geometric Brownian motion or the Black-Scholes equation is related to the
log-normal distribution \cite{Black73}.
Interrelations between the power-law distribution and the log-normal one have drawn
a great deal of attention in literature and the concept of the double Pareto distribution
has also been proposed \cite{Reed03, Mitzenmacher04}.
The Weibull distribution \cite{Weibull51} is also one of the widely applied distribution functions in a variety of systems \cite{Peterson85, Mcdowell96, Papadopoulou06, Jo07}. Emergence of the Weibull distribution has been attributed to diverse origins on the case-by-case basis, including the extreme value statistics \cite{Beirlant04, Bertin06}, fractal cracking \cite{Peterson85}, Fickian diffusion in the fractal and non-fractal spaces \cite{Kosmidis03PR, Kosmidis03JCP}, and the branching process \cite{Jo10}. Interrelations between the Weibull distribution and the log-normal one have also been studied and the similarity between them has been elucidated \cite{Brown95}.
There are also works which analyzed
a variety of time series data, e.g., from respiratory dynamics, electroencephalogram (EEG), heartbeats, and stock price records, and probed the asymmetry of distributions~\cite{Ohashi03, Braun98, Ehlers98, Sammon97}.

Even though some relations between these skew distributions are available, there still lacks theoretical understanding of how the skew distributions appear in evolving systems. Recently, a general framework has been proposed, based on the master equation describing evolving systems, to elucidate the emergence of skew distributions \cite{MYChoi09}.
Following this approach, we begin with the same master equation, to obtain the time evolution equation for the size distribution in evolving systems, which provides refinements and extensions. It is found that such skew distributions indeed emerge, depending on the growth, division, and production. In particular, the detailed conditions for power-law, log-normal, and Weibull distributions are clarified. In addition, numerical simulations are also carried out extensively; results are shown to agree well with the analytical ones.

There are five sections in this paper: In Sec. II we introduce the master equation approach to the proliferation process, including growth, division, and production. Section III is devoted to solving the evolution equation for various cases. Such skew distributions as power-law, log-normal, and Weibull distributions are obtained in the systems with uniform production, self-production, and binary fission. Section IV presents the results of numerical simulations, which confirm the analytical results in Sec. III. Finally, a summary together with a discussion of the results is given in Sec. V.

\section{Master equation}

We consider a system of $N$ elements, the $i$th of which is characterized by the size $x_i$
($i = 1, \ldots, N$).
The configuration of the system is specified by the sizes of all elements, $\{x_1^{}, \dots, x_N^{}\}$.  The probability $P(x_1^{}, \dots, x_N^{}; t)$ for the system to be in configuration $\{x_1^{}, \dots, x_N^{}\}$ at time $t$ is governed by the master equation
\begin{widetext}
  \begin{equation}
    \label{eq: master equation}
      \td{} P(x_1^{},\dots,x_{N}^{};t) = \sum_{i=1}^{N} \int d x_i'\, \bs{\omega(x_i' \rightarrow x_i^{}) P(x_1^{},\dots,x_i',\dots,x_{N}^{};t) - \omega(x_i \rightarrow x_i') P(x_1^{},\dots,x_{N}^{};t)} ,
  \end{equation}
\end{widetext}
where $\omega(x_i \rightarrow x_i')$ is the transition rate for the $i$th element to change its size from $x_i$ to $x_i'$. Here we consider the case that the size changes by the amount proportional to the current size.  The transition rate then takes the form
\begin{equation}
  \label{eq: transition rate}
  \omega(x \rightarrow x') = \lambda \delta(x'-(1{+}b)x)
\end{equation}
with the (mean) growth rate $\lambda$ and the growth factor $b$ \cite{MYChoi09}.

We are interested in the size distribution of the system $f(x,t)$, related to the probability $P(x_1^{}, \dots, x_N^{}; t)$ via:
\begin{equation}
  \label{eq: size distribution}
  f(x,t) = \frac{1}{N} \int d^N\!x\, \sum_{i=1}^{N} \delta(x_i - x) P(x_1^{},\dots,x_{N}^{}; t)\,,
\end{equation}
where $\int d^N\!x \equiv \int_0^\infty d x_{1}^{} \cdots \int_0^\infty d x_{N}^{}$.
The evolution equation for the size distribution $f(x,t)$ can be obtained straightforwardly from Eq. \eqref{eq: master equation}.
In the case that the total number of elements remains constant, the evolution equation for $f(x,t)$
reads
\begin{equation}
  \label{eq: time evolution equation for number conservation}
    \pd{}{t} f(x,t) = - \lambda f(x,t) + \frac{\lambda}{1+b} f\bp{\frac{x}{1{+}b},t}\,.
\end{equation}
Note the scale invariance of Eq.~\eqref{eq: time evolution equation for number conservation}: Changing the scale according to $x \rightarrow cx$ and multiplying Eq.~\eqref{eq: time evolution equation for number conservation} by $c$ for normalization, we obtain Eq.~\eqref{eq: time evolution equation for number conservation} in the form
\begin{equation}
\frac{\partial}{\partial t}cf(cx,t) = -\lambda cf(cx,t) +\lambda \frac{c}{1+b}f\left(\frac{cx}{1{+}b},t \right).
\end{equation}
This manifests that for $f(x,t)$ satisfying
Eq.~\eqref{eq: time evolution equation for number conservation}, $cf(cx,t)$ also satisfies
Eq.~\eqref{eq: time evolution equation for number conservation}, thus providing a solution.

On the other hand, when the total number of elements varies with time, i.e., $N=N(t)$,
it is not allowed to interchange the order of $d/dt$ and $\int d^{N}\!x$.
To circumvent this difficulty, we deal with the time derivative of $N(t)f(x,t)$:
\begin{widetext}
  \begin{eqnarray}
    \label{eq: time derivative of Nf}
      \pd{}{t} [N(t)f(x,t)] &=& \lim_{\Delta t \rightarrow 0} \frac{1}{\Delta t}
      \left[\int d^{N{+}\Delta N}\!x \sum_{i=1}^{N{+}\Delta N} \delta(x_i - x)
            P(x_1^{},\dots,x_{N{+}\Delta N}^{}; t{+}\Delta t) 
       - \int d^N\!x \sum_{i=1}^{N} \delta(x_i - x) P(x_1^{},\dots,x_{N}^{};t)\right] \nonumber \\
       &=& \lim_{\Delta t \rightarrow 0} \frac{1}{\Delta t} \left\{\int d^N\! x \sum_{i=1}^{N} \delta(x_i^{} - x) \left[P(x_1^{},\dots,x_{N}^{};t{+}\Delta t) - P(x_1^{},\dots,x_{N}^{};t)\right] \right. \nonumber \\
       & & \hspace{2cm} + \left.  \int d x_{N{+}1}^{} \cdots d x_{N{+}\Delta N}^{}
        \sum_{i=N{+}1}^{N{+}\Delta N} \delta(x_i^{} - x)
               P(x_{N{+}1}^{},\dots,x_{N{+}\Delta N}^{};t{+}\Delta t)\right\},
  \end{eqnarray}
where it has been noted that $\int dx_{N{+}1}^{} \cdots dx_{N{+}\Delta N}^{}\, P(x_1^{},\dots,x_{N{+}\Delta N}^{};t{+}\Delta t) = P(x_1^{},\dots,x_{N}^{}; t{+}\Delta t)$ and $\int d^{N}\!x\, P(x_1^{},\dots,x_{N{+}\Delta N}^{}; t{+}\Delta t) = P(x_{N{+}1}^{},\dots,x_{N{+}\Delta N}^{}; t{+}\Delta t)$.
\end{widetext}
The first term on the right-hand side (the second line) of Eq. \eqref{eq: time derivative of Nf} contains the information only on the existing elements (up to $N$) and is exactly the same as the case for fixed $N$, thus resulting in
\begin{equation}
  \label{eq: 1st term of Nf}
  (\textrm{1st term}) = N(t) \bs{-\lambda f(x,t) + \frac{\lambda}{1+b} f\bp{\frac{x}{1{+}b},t}} .
\end{equation}
The second term (the third line) of Eq. \eqref{eq: time derivative of Nf} is involved with the information on the newly produced elements and takes the form:
\begin{equation}
  \label{eq: 2nd term of Nf}
  (\textrm{2nd term}) 
  = \dot{N}(t) g(x,t) = rN(t) g(x,t) ,
\end{equation}
where $g(x,t)$, define to be
\begin{widetext}
\begin{equation}
  \label{eq: production distribution}
    g(x,t) \equiv \frac{1}{\Delta N} \int d x_{N{+}1}^{} \cdots d x_{N{+}\Delta N}^{}
       \sum_{i=N{+}1}^{N{+}\Delta N} \delta(x_i - x) P(x_{N{+}1}^{},\dots,x_{N{+}\Delta N}^{};t) ,
\end{equation}
\end{widetext}
corresponds to the size distribution of newly produced elements at time $t$
and $r$ is the mean production rate.
Namely, it is assumed that each element tends to produce a new one with rate $r$,
which in turn leads the total number of elements to increase in proportion to the current number:
$  \dot{N} = rN $.
Here care needs to be given to the limit $\Delta t \rightarrow 0$ since
$\Delta N$ is discrete, taking only integer values.
The expression $\Delta N = rN\Delta t$ is valid as long as $N\Delta t$ is finite,
so that $\Delta N$ remains discrete.
Accordingly, $\Delta t$ can be reduced down to $(rN)^{-1}$, which corresponds to the time interval
required to produce a new element ($\Delta N =1$).
This vanishes as $N$ grows large, thus allowing
$\lim_{\Delta t \rightarrow 0} \Delta N/\Delta t = \dot{N}$, etc.
With the help of Eqs. \eqref{eq: time derivative of Nf}-\eqref{eq: production distribution} and
\begin{equation}
  \label{eq: identity}
  \pd{}{t} f(x,t) = \frac{1}{N} \pd{}{t} \bs{N f(x,t)} - \frac{\dot{N}}{N} f(x,t)\,,
\end{equation}
the time evolution equation for the size distribution finally obtains the form:
\begin{equation}
  \label{eq: final version of time evolution equation}
  \pd{f(x,t)}{t} = -(r+\lambda) f(x,t) + \frac{\lambda}{1+b} f\bp{\frac{x}{1{+}b},t} + rg(x,t)\,.
\end{equation}

\section{Asymptotic solutions}

In this section we probe analytically the solutions of the evolution equations \eqref{eq: time evolution equation for number conservation} and \eqref{eq: final version of time evolution equation}.
In the case of no production ($r=0$), the time evolution of $f(x,t)$ is described by Eq. \eqref{eq: time evolution equation for number conservation}. Changing the variable from $x$ to $\ln x \equiv X$,
we write the distribution function for $X$: $F(X,t) = e^X f(e^X,t)$.  Correspondingly, Eq. \eqref{eq: time evolution equation for number conservation} becomes
\begin{equation}
  \label{eq: no production (modified)}
  \pd{}{t} F(X,t) = -\lambda F(X,t) + \lambda F(X{-}a,t),
\end{equation}
with $a \equiv \ln (1{+}b)$. Given the initial distribution $f(x,0) = \delta (x-x_0)$ or equivalently, $F(X,0) = \delta (X-X_0)$, $F(X,t)$ vanishes unless $X =X_n \equiv X_0 + an$ for any integer $n$
(note that this corresponds simply to $x_n = x_0 e^{an} = (1+b)^n x_0$).
With the definition $F_n^{}(t) \equiv F(X_0{+}an,t)$, Eq. \eqref{eq: no production (modified)} reads
\begin{equation}
  \label{eq: no production (discrete)}
  \dot{F}_n^{}(t) = -\lambda F_n^{}(t) + \lambda F_{n-1}^{} (t) ,
\end{equation}
which bears the Poisson distribution
\begin{equation}
  \label{eq: Poisson's distribution}
  F_n^{}(t) = \frac{1}{n!} \bp{\lambda t}^n e^{-\lambda t}
\end{equation}
for $n \ge 0$ and $F_n^{}(t) = 0$ otherwise \cite{Risken}.

This result can be also recovered by considering the Fourier transform
$\tilde F(k,t)$ of Eq. (\ref{eq: no production (discrete)}), which
satisfies
\bb\label{eq: FT}
\partial_t \tilde F(k,t)=\lambda (e^{-ika}-1)\tilde F(k,t).
\ee
The solution $ \tilde F(k,t)=\exp[-\lambda(1-e^{-ika})t]\tilde F(k,0)$ of Eq. (\ref{eq: FT})
leads directly to
\bb\label{eq: no prod general}
F(X,t)&=&\int dX' G_F(X,X';t) F(X',0) \nn \\
      &=&G_F(X,X_0;t),
\ee
where
$G_F(X,X';t) = (2\pi)^{-1} \int dk \exp[ik(X-X')-\lambda (1-e^{-ika})t)]$
and the initial condition $F(X',0) = \delta (X'-X_0)$ has been noted.

When $t=0$, we have simply $F(X,0) = G_F(X,X_0;t=0)=\delta(X-X_0)$.
For large values of $\lambda t$, the function $G_F$ is dominated by small values of
$k$; in this limit, using the saddle-point analysis, we find that
Eq. \eqref{eq: Poisson's distribution} is equivalent to the normal distribution:
\begin{equation}
  \label{eq: normal distribution}
  F(X, t| X_0) = \frac{1}{\sqrt{2 \pi} \sigma_t^{}} \exp\bs{-\frac{\bp{X-X_{0}-\mu_t^{}}^2}{2 \sigma_t^2}} ,
\end{equation}
where $\mu_t^{} = a\lambda t$, $\sigma_t^{} = a\sqrt{\lambda t}$,
and $F(X, t| X_{0})$ stands for the distribution $F(X,t)$ subject to the initial distribution
$F(X,0) = \delta(X-X_{0})$. The same result is obtained when the growth factor $b$
and hence $a$ are small; this allows us to expand the term $\exp(-ika)$
in the exponential integrand and to perform the gaussian integration.
We may also expand $G_F$ directly as the series
\bb\nn
G_F(X,X';t)&=&\sum_{n\ge 0}\int\frac{dk}{2\pi}e^{
ik(X-X')-\lambda t}\frac{(\lambda t)^n}{n!}e^{-ikan}
\\ \nn
&=&\sum_{n\ge 0}\frac{(\lambda t)^n}{n!}e^{-\lambda t}\delta(X-X'-an)
\\
&=&\sum_{n\ge 0}F_n(t)\delta(X-X'-an),
\ee
which is nothing but the expansion in terms of the Poisson distribution at given time.

In consequence, the original size distribution assumes the log-normal
distribution in the long-time limit:
\begin{equation}
  \label{eq: log-normal distribution}
  f(x, t| x_{0}) =\frac{1}{\sqrt{2 \pi} \sigma_t^{}x} \exp \bs{-\frac{\bp{\ln x - \ln x_0 -\mu_t^{}}^2}{2 \sigma_t^2}} \,,
\end{equation}
where $f(x,t| x_{0})$ denotes the size distribution $f(x,t)$ subject to the initial distribution $f(x,0) = \delta(x-x_{0})$. Given an arbitrary initial distribution $f(x,0)$, the solution thus takes the form
\bb \label{eq: no production solution}
  f(x,t)=\int d x_{}'G_f(x,x';t)f(x_{}',0)
\ee
with
\bb \label{Gf}
G_f(x,x';t)=\frac{1}{x}G_F(\ln x, \ln x';
t)\,.
\ee
Equation \eqref{eq: no production solution}, like Eq. \eqref{eq: time evolution equation for number conservation}, is invariant under the scale transformation $x \rightarrow cx$ and $x_{}' \rightarrow cx_{}'$.

We now consider the case $r\neq 0$.
Expanding the second term on the right-hand side of Eq. \eqref{eq: final version of time evolution equation}: \begin{equation}
  \label{eq: Taylor expansion}
  f\bp{\frac{x}{1{+}b},t} = \sum_{n=0}^{\infty}\frac{(-1)^n}{n!}\bp{\frac{b}{1+b}}^n x^n \frac{\partial ^n}{\partial x^n} f(x,t)\,,
\end{equation}
we write Eq. \eqref{eq: final version of time evolution equation} in the compact form
\begin{equation}
\mathcal{L} f(x,t) = rg(x,t),
\end{equation}
where the linear operator $\mathcal{L}$ is given by
\begin{equation}
  \label{eq: linear operator}
  \mathcal{L} \equiv \pd{}{t} + r + \lambda - \frac{\lambda}{1+b}\sum_{n=0}^{\infty}\frac{(-1)^n}{n!}\bp{\frac{b}{1+b}}^n x^n \frac{\partial ^n}{\partial x^n}\,.
\end{equation}
Since the solution for $r=0$, Eq. \eqref{eq: log-normal distribution}, reduces to the Dirac delta function $\delta(x-x_{0})$ in the limit $t \rightarrow 0$, Green's function $G(x,x';t,t')=G(x,x';t{-}t')$ for $\mathcal{L}$ can be written in the form
\begin{equation}
  \label{eq: Green's function}
  G\bp{x,x';t{-}t'} = G_f\bp{x,x';t{-}t'} e^{-r\bp{t-t'}} \theta\bp{t-t'}\,,
\end{equation}
with the Heaviside step function $\theta(t-t')$.

This leads to the general solution of Eq. \eqref{eq: final version of time evolution equation} in the form:
\begin{widetext}
  \begin{equation}
    \label{eq: general solution}
      f(x,t) =
      e^{-rt} \int d x_{}'\, G_f(x,x';t)\, f(x_{}',0)
      + r \int_0^t dt'\, e^{-r(t-t')} \int dx'\, G_f(x,x';t-t') g(x', t') ,
  \end{equation}
\end{widetext}
where Green's function reduces to the log-normal distribution in the long-time limit
($t\rightarrow \infty$):
\bb \nn
G_f(x,x';t) \longrightarrow \frac{1}{\sqrt{2 \pi} \sigma_t^{}x}
              \exp\left[-\frac{(\ln x - \ln x_{}' -\mu_t^{})^2}{2 \sigma_t^2}\right]\,.
\ee
The first term on the right-hand side of Eq. \eqref{eq: general solution}, which describes the information on the initial distribution, decreases exponentially in time; the second term thus becomes dominant in the long time.

\subsection{Uniform-size production}
In case that new elements are produced in uniform size $x_0$, i.e.,
$g(x,t) = \delta (x-x_0)$, the asymptotic solution in the long-time limit reads
\begin{widetext}
\begin{eqnarray}
  \label{eq: uniform size production stationary solution}
f_s^{}(x) \equiv f(x,t {\rightarrow} \infty) = \lim_{t{\rightarrow}\infty}
r\int_0^t dt'\, e^{-r(t-t')}G_f(x,x_0;t-t')
=\frac{r}{x}\int \frac{dk}{2\pi}\,\frac{e^{ik\ln(x/x_0)}}{r+\lambda(1-e^{-ika})} .
\end{eqnarray}
\end{widetext}
The integral over $k$ can be performed with the help of the decomposition $ka= 2\pi n-\theta$,
where $n$ is an integer and $0\le \theta < 2\pi$.
Equation (\ref{eq: uniform size production stationary solution}) then obtains the form of
an infinite sum over integers $n$:
\bb \nn
f_s^{}(x)=\frac{r}{ax}\sum_n e^{2i(\pi/a) n\ln(x/x_0)}
          \int_0^{2\pi}\frac{d\theta}{2\pi}
                   \frac{e^{-(i/a)\theta \ln(x/x_0)}}{r+\lambda-\lambda e^{i\theta}}.
\ee
We notice that the integral over $\theta$ is independent of $n$, and use the Dirac comb identity
to write
\bb\nn
f_s^{}(x)&=&\frac{r}{ax}\sum_n \delta \left(n- a^{-1}\ln(x/x_0)\right)
          \int_0^{2\pi}\frac{d\theta}{2\pi}
                    \frac{e^{-(i/a)\theta\ln(x/x_0)}}{r+\lambda-\lambda e^{i\theta}} \\
        &=&\frac{r}{ax}\sum_n \delta\left(n-a^{-1}\ln(x/x_0)\right)
           \int_0^{2\pi}\frac{d\theta}{2\pi}
                   \frac{e^{-in\theta}}{r+\lambda-\lambda e^{i\theta}}, \nn
\ee
where it has been noted that the argument $a^{-1}\ln(x/x_0)$ is constrained by each integer $n$.
The integral over $\theta$ can be transformed into one over a unit circle in a complex
plane:
\bb\nn
\int_0^{2\pi}\frac{d\theta}{2\pi}
\frac{e^{-in\theta}}{r+\lambda-\lambda e^{i\theta}}
=\oint\frac{dz}{2i\pi z}\frac{z^{-n}}{r+\lambda-\lambda z},
\ee
where we have set $e^{i\theta} \equiv z$.
This integral vanishes when $n$ is negative, i.e., for $x<x_0$;
otherwise Cauchy's residue theorem yields
\bb\nn
\oint\frac{dz}{2i\pi z}\frac{z^{-n}}{r+\lambda-\lambda z}=
\frac{\lambda^n}{(r+\lambda)^{n+1}}
\ee
for $n\ge 0$.
Setting finally $x_n=x_0e^{an} =x_0 (1+b)^n$, we obtain the stationary distribution
\bb\label{eq: uniform size production solution large}
f_s(x)=\frac{r}{r+\lambda}\sum_{n\ge 0}
\left ( \frac{x}{x_0} \right )^{-\frac{\ln (1+r/\lambda)}{\ln(1+b)}}\delta(x-x_n).
\ee

On the other hand, the stationary solution can be obtained simply by putting
$\partial f /\partial t = 0$ and $g(x,t) = \delta (x-x_0)$ in Eq. \eqref{eq: final version of time evolution equation}.
In this way, we still obtain the power-law distribution for $x >x_0$:
\begin{equation}
 \label{eq: stationary power}
 f_s (x) \sim x^{-\alpha}
\end{equation}
with the exponent
$\alpha = 1 + \frac{\ln{(1{+}r/\lambda)}}{\ln{(1{+}b)}}$,
which should be exact in the stationary configuration.
Equations~\eqref{eq: uniform size production solution large}
and \eqref{eq: stationary power} give the same kind of power-law distribution in the
long-time limit.
It is indeed possible to recover from Eq. \eqref{eq: uniform size
production solution large} the simple stationary power-law
solution for small $b$ or $a$. In such a continuous-growth limit,
we also have $r\rightarrow 0$, with the ratio $r/b$ and parameter
$\lambda$ kept finite, since $[\ln(1{+}b)]^{-1}\ln
(1{+}r/\lambda)$ and thus $\alpha$ would diverge otherwise. In
this case Eq. \eqref{eq: uniform size production solution large}
may be approximated by replacing the discrete sum over $n$ by an
integral over the variable $u=an$ which becomes continuous in the
limit $a\rightarrow 0$: \bb\nn
f_s(x)&=&\frac{r}{r+\lambda}\frac{1}{x}\sum_{n\ge 0} \left (
\frac{x}{x_0} \right )^{-\frac{\ln
(1+r/\lambda)}{\ln(1+b)}}\delta(an-\ln(x/x_0))
\\ \nn
&\approx &\frac{r}{r+\lambda}\frac{1}{ax}\int_0^{\infty} d\,u
\left ( \frac{x}{x_0} \right )^{1-\alpha} \delta (u-\ln(x/x_0))
\\
&=&\frac{r}{\lambda bx_0}
\left ( \frac{x}{x_0} \right )^{-\alpha} \theta (x-x_0) ,
\ee
where $\theta$ is the Heaviside step function. This normalized power-law distribution has
therefore the same exponent $\alpha = 1 + [\ln(1{+}b)]^{-1}\ln (1{+}r/\lambda) \approx 1+ r/b\lambda$
as the stationary one given by Eq. \eqref{eq: stationary power} in the continuous-growth limit.
Accordingly, our general solution given by Eq. \eqref{eq: general solution} 
demonstrates that the distribution, starting from the Dirac delta distribution, evolves to the
power-law distribution in the uniform-size production process.

\subsection{Self-size production}
When an element produces a new element of the same size as itself, the size distribution of newly produced elements is identical to that of the system, i.e., $g(x,t)=f(x,t)$.
In this case, the size distribution of newly produced elements is not given externally, and Eq. \eqref{eq: general solution} no longer provides the solution of Eq. \eqref{eq: final version of time evolution equation} but makes an integral equation for $f(x,t)$.
Remarkably, however, putting $g(x,t)=f(x,t)$ in Eq. \eqref{eq: final version of time evolution equation} reduces the equation precisely to Eq. \eqref{eq: time evolution equation for number conservation}.
Namely, the case of self-size production is effectively the same as the case of no production ($r=0$), except for the increase of the total number of elements due to the production of new elements. In consequence, the size distribution of the system with self-size production is given by Eq. \eqref{eq: no production solution}. Starting from the initial configuration that all elements have the identical size $x_{0}$, the size distribution develops to the log-normal distribution:
\begin{equation}
  \label{eq: self-size production solution}
  f(x,t) = \frac{1}{\sqrt{2 \pi} \sigma_t^{} x}\exp\bs{-\frac{\bp{\ln x -\mu_t^{}}^2}{2 \sigma_t^2}}
\end{equation}
with $\mu_t^{} =\ln(1{+}b) \lambda t + \ln x_{0}$ and
$\sigma_t^{} = \ln(1{+}b) \sqrt{\lambda t}$.

\subsection{Binary fission}
The binary fission by means of division in size may be interpreted as follows: The size of an element is reduced by the multiplicative factor $\beta \,< 1$ at rate $r$ and a new element with the size reduced by the multiplicative factor $1-\beta$ is produced at the same rate $r$.
As a result, the corresponding evolution equation obtains the form
\begin{widetext}
\begin{equation}
  \label{eq: binary fission}
    \pd{f(x,t)}{t} = -(2r +\lambda) f(x,t) + \frac{\lambda}{1+b} f\bp{\frac{x}{1{+}b},t}
    + \frac{r}{\beta} f\bp{\frac{x}{\beta},t} + \frac{r}{1-\beta} f\bp{\frac{x}{1{-}\beta},t} ,
\end{equation}
\end{widetext}
which describes both growth and reduction of the size due to binary fission.
Although it can be easily extended further to describe more complicated processes,
we consider only one each kind of size-growth process and size-reduction process,
which suffices for our study here.
Specifically, we suppose that the size of an element remains constant while the element gives birth to an offspring of the size proportional to its size with the multiplicative factor $1-\beta$:
\begin{widetext}
\begin{equation}
  \label{eq: binary fission one size-reduction process}
    \pd{f(x,t)}{t} = -(r +\lambda) f(x,t) + \frac{\lambda}{1+b} f\bp{\frac{x}{1{+}b},t}
    + \frac{r}{1-\beta} f\bp{\frac{x}{1{-}\beta},t}\,,
\end{equation}
\end{widetext}
with $b > 0$ and $0 < \beta < 1$. Note that, in the half-size division process ($\beta=1/2$), Eqs. \eqref{eq: binary fission} and \eqref{eq: binary fission one size-reduction process} become identical.

In contrast to the cases of uniform-size production and self-size production, the size distribution of newly produced elements depends on that of the system and does not cancel out, respectively.
Making use of the scaling symmetry of Eq.~\eqref{eq: binary fission one size-reduction process},
one is tempted to take a log-normal or a power-law distribution as an ansatz solution.
However, neither the log-normal distribution nor the power-law distribution can make a stable solution,
as manifested easily by substitution.
This may be inferred from the observation that the linear operator $\mathcal{L}$ for
Eq.~\eqref{eq: binary fission one size-reduction process}
cannot be cast into the same form as that for Eq.~\eqref{eq: linear operator}.
Therefore we introduce another distribution observing the scaling symmetry,
namely, the Weibull distribution:
\begin{equation}
  \label{eq: Weibull distribution}
  f(x,t) = \frac{\gamma}{x}\bp{\frac{x}{\eta}}^{\gamma} \exp \bs{-\bp{\frac{x}{\eta}}^\gamma}
\end{equation}
with the shape parameter $\gamma$ and the scale parameter $\eta$.
There exist other distribution functions with the same symmetry, for example, the
generalized exponential distribution and the gamma distribution;
they also fail to provide asymptotic solutions of Eq.~\eqref{eq: binary fission one size-reduction process}.

Henceforth we denote $y \equiv \ln (x/\eta)$, $u \equiv 1-(1+b)^{-\gamma}$, and $v \equiv 1-(1-\beta)^{-\gamma}$ for convenience.
Expanding $(x/\eta)^\gamma$ at $\gamma = 0$,
\begin{equation}
  \label{eq: expansion with respect to gamma}
  \bp{\frac{x}{\eta}}^\gamma = \sum_{k=0}^{\infty} \frac{1}{k!} (\gamma y)^k\,,
\end{equation}
we write the time derivative of the Weibull distribution in the form
\begin{equation}
  \label{eq: time derivative modified}
  \frac{1}{f} \pd{f}{t} = \frac{\dot{\gamma}}{\gamma} + \bp{\dot{\eta}\frac{\gamma}{\eta} - \dot{\gamma}y} \sum_{k=1}^{\infty} \frac{1}{k!} (\gamma y)^k \,,
\end{equation}
while the right-hand side of Eq. \eqref{eq: binary fission one size-reduction process} divided by $f(x,t)$ becomes
\begin{widetext}
  \begin{equation}
    \label{eq: size part}
      \frac{1}{f} (\textrm{rhs})
      = -(\lambda + r) + \lambda (1-v) e^v + r (1-u) e^u + \lambda (1-v) e^v \sum_{k=1}^{\infty} \frac{1}{k!} B_k^{}(v)(\gamma y)^k + r (1-u) e^u \sum_{k=1}^{\infty} \frac{1}{k!} B_k^{}(u)(\gamma y)^k
  \end{equation}
\end{widetext}
with the Bell polynomial given by \cite{Roman}
\begin{equation}
  \label{eq: Bell polynomial}
  B_k^{}(z) \equiv e^{-z} \sum_{n=0}^{\infty} \frac{n^k}{n!}z^n .
\end{equation}

Comparison of Eqs. \eqref{eq: time derivative modified} and \eqref{eq: size part} gives, for small $\gamma$,
\begin{equation}
  \label{eq: gamma}
  \dot{\gamma}= -\frac{\gamma^3}{2}\bc{\lambda \bs{\ln (1{+}b)}^2 + r \bs{\ln (1{-}\beta)}^2} + O(\gamma^4)
\end{equation}
to the zeroth order in $y$, and
\begin{equation}
  \label{eq: eta}
  \dot{\eta} = \eta\bs{\lambda \ln (1{+}b) + r \ln (1{-}\beta )} + O(\gamma)\,,
\end{equation}
to the $n$th order in $y$ for all $n \geq 1$,
where the expansion of $v$ about $\gamma=0$,
\begin{equation}
  \label{eq: expansion of v}
  v \equiv 1-\bp{\frac{1}{1+b}}^\gamma = - \sum_{k=1}^{\infty} \frac{(-1)^k}{k!}\bs{\gamma \ln (1{+}b)}^k ,
\end{equation}
and similar expansion of $u$ (with $-\beta$ in the place of $b$) have been used.
To the leading order, the solutions of Eqs. \eqref{eq: gamma} and \eqref{eq: eta} under the initial conditions $\gamma (0)=\gamma_0^{}$ and $\eta (0)=\eta_0^{}$ read
\begin{subequations}
  \label{eq: solution of gamma and eta}
  \begin{align}
    \gamma(t) &= \frac{\gamma_0^{}}{\sqrt{1+ C\gamma_0^2 t}}
        \label{eq: solution of gamma}\\
    \eta(t) &= \eta_0^{}\, e^{Dt}\,, \label{eq: solution of eta}
  \end{align}
\end{subequations}
where $C \equiv \lambda \bs{\ln (1{+}b)}^2 + r \bs{\ln (1{-}\beta)}^2$ and
$D \equiv \lambda \ln (1{+}b) + r \ln (1{-}\beta)$.
For this to be valid, one should have sufficiently small $\gamma$ and $y$ (or large $\eta$). In the asymptotic limit the conditions of small $\gamma$ and large $\eta$
require both $C$ and $D$ to be positive. Note that while $C$ is always positive, $D$ is positive only for $r / \lambda < -\ln (1{+}b) / \ln (1{-}\beta)$.
%
In conclusion, for $r / \lambda < -\ln (1{+}b) / \ln (1{-}\beta)$, the Weibull distribution in Eq. \eqref{eq: Weibull distribution} is asymptotically exact, where the shape parameter $\gamma$ decreases algebraically in time and the scale parameter $\eta$ grows exponentially in time.

\subsection{Autocorrelations}
In the case of an asymmetric distribution of variables, there can exist power-law correlations
in the variables for suitably adjusted parameters~\cite{Podobnik05}.
To probe the possibility of long-time correlations in the evolution governed by
Eq. \eqref{eq: final version of time evolution equation}, we consider the autocorrelation function
at stationarity:
\begin{equation}
\label{auto}
C(t) \equiv \frac{\langle x(\tau) x(\tau{+}t) \rangle - \langle x(\tau)\rangle \langle x(\tau{+}t) \rangle}
{\sqrt{ \langle x^2(\tau) \rangle - \langle x(\tau) \rangle^2 }
\sqrt{ \langle  x^2(\tau{+}t) \rangle - \langle x(\tau{+}t) \rangle^2}}
\end{equation}
where $\langle x(t)\rangle \equiv \int dx\, f(x,t) x$
and 
$\langle {x(\tau)x(\tau{+}t)} \rangle \equiv \int dx\,dx' \,G
(x,x'; t)f(x',\tau) x x'$
with appropriate Green's function $G(x,x'; t)$ and $\tau$ being
large enough for the system to be in the stationary state.

It is straightforward to compute Eq. \eqref{auto}, which turns out
to exhibit exponentially decaying behavior in the long-time limit.
Specifically, in the case of the self-size production or no
production ($r=0$), Green's function reduces simply to Eq.
(\ref{Gf}): $G(x,x'; t) = G_f (x,x'; t)$, and the resulting
autocorrelation function reads
\begin{equation}
C(t)
\sim \exp{\left(-\frac{b^2 \lambda}{2}t\right)}.
\end{equation}
For the branching process, Green's function is given by
\begin{widetext}
\begin{equation}
G(x,x'; t) = e^{-r t}G_f (x,x'; t)
+ r\int_{0}^{ t} dt^\prime e^{-r(t-t^\prime)}
G_f (x, (1{-}\beta)x'; t{-}t^\prime ),
\end{equation}
\end{widetext}
which leads to
\begin{equation}
C(t) \sim \exp{\left( -\frac{b^2
\lambda +\beta^2 r}{2} t\right)}.
\end{equation}
Finally, when there is a source producing new elements of uniform
size, Green's function takes the form
\begin{widetext}
\begin{equation}
G(x,x';t) = e^{-r t} G_f (x,x'; t)
+ r\int_{0}^{t} dt^\prime e^{-r(t-t^\prime)}
G_f (x,x_0 ; t{-}t^\prime)
\end{equation}
\end{widetext}
where $x_0$ is the size of newly produced elements. In this case
the autocorrelation function depends on the production rate $r$:
For $r<2b\lambda +b^2 \lambda$, it reads
\begin{equation}
C(t) \sim
\exp{\left(-\frac{r+b^2 \lambda}{2} t \right)}
\end{equation}
whereas for $r\geq 2b\lambda +b^2 \lambda$, we have
\begin{equation}
C(t) \sim
\exp{\left[ (-r+b \lambda ) t \right]}.
\end{equation}

In all cases, the autocorrelations decay exponentially in time. We
have also carried out numerical simulations to compute
autocorrelation functions, to obtain results in full agreement
with the analytical results listed above. It is thus concluded
that long-time autocorrelations do not exist in the growth process
considered here. This apparently implies that skew distributions
do not necessarily result from long-time autocorrelations of the
data variables.

\section{Numerical Simulations}

We carry out numerical simulations to check the analytical results in the preceding section.
The master equation \eqref{eq: master equation} is integrated directly to give the probability distribution for no-production, uniform-size production, self-size production, and half-size division processes. At each time step given by an integer multiple of $\Delta t$ in simulations, a trial move is attempted according to the growth probability
$\Lambda \equiv \lambda \Delta t$ and the production probability $R \equiv r \Delta t$.
To access the continuous-time case, $\Delta t$ should be taken to be sufficiently small.
Here we choose $\Delta t = 0.001$ and perform simulations of $10^2$ different samples,
over which obtained results are averaged. Initially, all elements in each sample are taken to be of the unit size
($x_0 =1$) and the number of elements is usually adjusted so that each sample in the final data consists of
$N=10^4$ to $10^6$ elements. 
These simulation conditions have been varied, to give no appreciable differences.

\begin{figure}
\centering
\includegraphics[width=0.31\textwidth, angle=270]{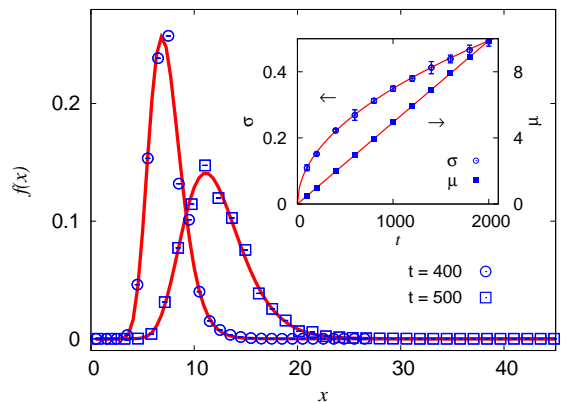}
\caption{\label{f:no production}
Evolution of the size distribution in the case of no production, $r=0$.
Circles and squares plot simulation data for $\lambda = 0.2$ and $b=0.025$
at time $t=400$ and $500$, respectively.
Solid lines represent the corresponding analytic solutions,
which are log-normal distributions with $(\mu, \sigma) = (1.975, 0.2209)$ and $(2.469, 0.2469)$, respectively.
The inset shows $\sigma$ (circles) and $\mu$ (squares) versus time $t$,
disclosing excellent agreement between simulation results (symbols) and analytical ones (solid lines).
Error bars have been estimated by the standard deviations of the simulation data, and are smaller
than the symbols.
}
\end{figure}

When no new elements are produced, the simulation results are shown in Fig.~\ref{f:no production}, and compared with the analytic solutions.
It is observed that the distributions obtained from simulations fit well to the log-normal distributions described by the analytical expression in Eq. \eqref{eq: no production solution}. The asymptotic solution also gives excellent estimate for the time evolution of the mean and standard deviation parameters (see the inset). This supports strongly the validity of the analysis in Sec. III, including the log-normal kernel in Eq. \eqref{eq: log-normal distribution} and the asymptotic general solution in Eq. \eqref{eq: general solution}.

\begin{figure}
\centering
\includegraphics[width=0.31\textwidth, angle=270]{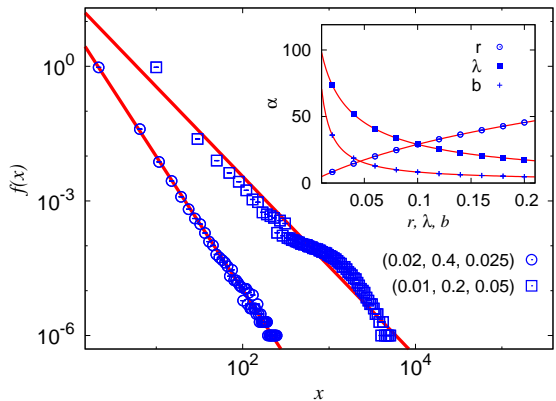}
\caption{\label{f:uniform size production}
Evolution of the size distribution in the case of uniform-size production.
Circles and squares plot simulation data for $(r, \lambda, b) = (0.02, 0.4, 0.025)$
at time $t=500$ and $(0.01, 0.2, 0.05)$ at $t=700$, respectively, in the log-log scale.
Solid lines represent stationary solutions, with the slope $-\alpha = -2.976$ and $-2$, respectively, given by Eq. \eqref{eq: stationary power}.
The inset shows the exponent $\alpha$ of the distribution versus various parameters:
Circles display $\alpha$ vs. the production rate $r$ for $\lambda=0.1$ and $b=0.025$;
squares present $\alpha$ vs. the growth rate $\lambda$ for $r=0.1$ and $b=0.025$,
and crosses $\alpha$ vs. the growth factor $b$ for $r=\lambda=0.1$.
Solid lines represent analytical solutions.
Error bars estimated by the standard deviations of the simulation data are smaller than the symbols.
}
\end{figure}

Figure~\ref{f:uniform size production} shows the simulation results for the case of uniform-size production,
where the initial number of elements has been adjusted in such a way that the system ends with $N=10^6$ elements
for both cases, $(r, \lambda, b) = (0.02, 0.4, 0.025)$ and $(0.01, 0.2, 0.05)$.
For comparison, analytical results are shown together, manifesting power-law behavior. In particular, the inset exhibits the exponent $\alpha$ of the power-law distribution versus $r$, $\lambda$, and $b$;
there simulation results are compared with the analytical solution
given by Eq.~\eqref{eq: uniform size production solution large} or Eq. \eqref{eq: stationary power}.
As expected, excellent agreement is observed.

For the self-size production process $g(x,t)=f(x,t)$, it is expected that simulations give the same results as the no-production cases. To confirm this, we carry out simulations of the system with the production rate $r=0.01$ and other parameters the same as those in Fig. \ref{f:no production} (i.e., $\lambda = 0.2$ and $b=0.025$).
The resulting distributions at $t=400$ and $500$ (not shown) indeed turn out identical to those in Fig. \ref{f:no production}.
This also implies that the contributions of the size distribution of newly produced elements to the change in the size distribution of the system are described correctly by Eq. \eqref{eq: final version of time evolution equation}.


\begin{figure}
\centering
\includegraphics[width=0.31\textwidth, angle=270]{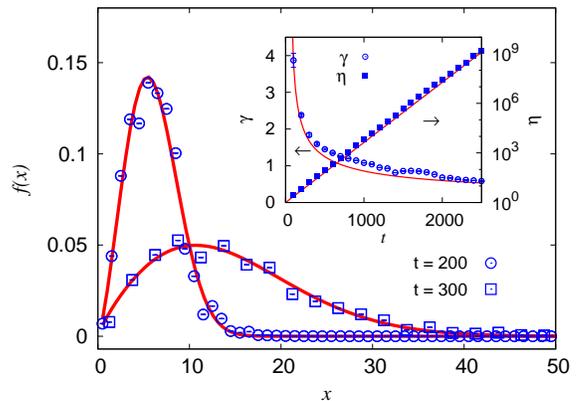}
\caption{\label{f:binary fission}
Evolution of the size distribution in the half-size division process with $r=0.002$, $\lambda=0.2$, and $b=0.05$.
Circles and squares plot simulation data at time $t=200$ and $300$, respectively.
Solid lines represent the fitted Weibull distribution
with $(\gamma, \eta) = (2.380, 6.834)$ for $t=200$ and $(1.842, 16.35)$ for $t=300$, respectively.
Error bars estimated by the standard deviations are not larger than the symbols.
The inset shows $\gamma$ (circles) and $\eta$ (squares) versus time $t$, disclosing good agreement between simulation results (symbols) and analytical ones (solid lines).
}
\end{figure}

Lastly, branching processes are simulated and the results for the half-size division process ($\beta=1/2$) are illustrated in Fig.~\ref{f:binary fission}. The resultant distributions fit well to the Weibull distribution. The shape parameter $\gamma$ and the scale parameter $\eta$ evolve in time, the agreement of which with the analytical results gets better as time goes on. This suggests that the branching process provides inherent mechanism of the Weibull distribution, the general origin of which has remained unresolved.

\section{Discussion}

We have established a model for describing the size distribution in evolving systems and probed its solutions in several cases both analytically and numerically.
Obtained are skew distributions including power-law, log-normal, and Weibull distributions, depending on the production rate, growth rate, and growth factor.
Specifically, power-law distributions are obtained when new elements of uniform size are produced; log-normal distributions emerge in case that each element tends to produce either a new one of the same size as itself or no element at all.
Further, branching processes give rise to Weibull distributions, suggestive of the origin of the Weibull distribution.

Emergence of the Weibull distribution in our model needs careful consideration: As the asymptotic expansion does not guarantee uniqueness of the solution, there may exist other solutions which are distinguishable from our solution by the asymptotic expansion to the leading order. Further, there still lacks understanding of the time evolution of the Weibull distribution, e.g., ranges of the production and growth rates in which the Weibull distribution provides quantitatively better fitting have not searched enough. The predicted time evolution of the shape and scale parameters is not exact, in contrast with the log-normal distribution for the self-size production process. Nevertheless, the physical meaning as to the emergence of the Weibull distribution is suggested convincingly and the accuracy of the asymptotic expansion is also supported to some degree by numerical simulations.

With the insight gained from this study, we may understand skew distributions observed in various systems. For instance, the power-law distribution of the populations in the U.S. cities \cite{Newman05} can be understood as an evolving system, where small towns are created and then grow to larger cities. The number of passengers in an intra-urban subway system \cite{KSLee08} may also be treated by this model:
The numbers of passengers making trips between two subway stations or using a subway station
are indeed observed to follow log-normal or Weibull distributions. 
The size distribution of beta cells in pancreatic islets, which has been experimentally measured and theoretically predicted \cite{Jo07}, is also described by a log-normal or Weibull distribution.
It is expected that there are many additional evolving systems with skew distributions, explained properly by this model.

It is remarkable that ubiquitous emergence of such representative skew distributions as
power-law, log-normal, and Weibull distributions is successfully described within a single framework.
Applications to real systems exhibiting these skew distributions and more detailed understanding of the emergence of the Weibull distribution are left for further study.

\begin{acknowledgments}
One of us (MYC) thanks Institut Jean Lamour, D\'epartement de Physique de la Mati\`ere et des Mat\'eriaux at Universit\'e Henri Poincar\'e, where part of this work was carried out, for hospitality during his stay.
This work was supported by the National Research Foundation of Korea through the Basic Science Research Program (Grant No. 2009-0080791).
\end{acknowledgments}

\bibliography{skew}

\end{document}